\documentclass[aps,prb,twocolumn,groupedaddress]{revtex4}

\usepackage[final]{graphicx}% Include figure files

\newcommand{\TAU}{\ensuremath{\tau{}}}
\newcommand{\NDIFF}{\ensuremath{n_{\mathrm{diff}}}}

\begin{document}

%\preprint{}

\title{ Emergence of Conduction Channels in Lithium Silicate}


\author{Heiko Lammert}
\email[]{hlammert@uni-muenster.de}
\author{Andreas Heuer}
\email[]{andheuer@uni-muenster.de}
%\thanks{}
\affiliation{Institute of Physical Chemistry and Sonderforschungsbereich 458,
Corrensstra{\ss}e 30, D-48149 M\"unster, Germany }


\date{\today}

\begin{abstract}

The existence of conduction channels in lithium silicate
$\mathrm(Li_2O)(SiO_2)$ is investigated. Regions of the system where
many different ions pass by form channels and are thus spatially
correlated. For a closer analysis the properties of the individual
ionic sites are elucidated. 
The mobility of ions in single sites is found
to depend strongly on the number of bridging oxygens in the
coordination shell. The channels are not reflected in the network
structure as obtained from the distribution of the bridging
oxygens. Spatial correlations similar to those found in
the silicate also emerge from studying the dynamics of particles 
in a simple random lattice model. 
This supports the suggestion that the observed spatial correlations
can be viewed in analogy to the emergence of percolation paths.

\end{abstract}

\pacs{61.20.Ja,61.43.Fs,66.30.Hs} 
%\keywords{}

\maketitle


\section{\label{sec_intro}Introduction}

The presence of conduction pathways has been proposed to
rationalize the fast conduction in ion conductors like alkali
silicates. Based on EXAFS data on the coordination environments in
these systems, Greaves proposed the modified random network model
for glasses\cite{greaves:1985}. It predicts the aggregation of
modifier ions into channels lined by non-bridging oxygen atoms.
Later NMR studies\cite{Yap:1995,gee:1996} reported both
inhomogeneous and homogeneous distributions of cations in the
structure with a clustering of alkali ions in the low
concentration regime. These results are backed by molecular
dynamics (MD) simulations of alkali silicate where also only for
low alkali concentrations a mild clustering of alkali ions has
been found \cite{Heuer:2002,horbach:2001}.

In his analysis of MD simulations of sodium silicate, Jund et al
\cite{jund:2001} focussed on the most mobile cations. They
introduced \NDIFF{}, the number of different ions visiting a given
subvolume, to determine those parts of the system taking part in
effective long range transport. They report that the regions of
the silicate network, visited by the largest number of {\it
different} ions form a network of blobs and connecting channels
(denoted {\it conduction channels}). Stated differently, these
regions show a significant spatial correlation beyond the
nearest-neighbor shell. In particular, these spatial correlations
are much stronger than any possible correlation of alkali
positions themselves. Thus these conduction channels do not simply
emerge from clustering of alkalis\cite{jund:2001} .

Interestingly, all static structure factors for sodium-silicate
exhibit a prepeak at $q_1 \approx 0.95 $ \AA$^{-1}$,
corresponding to the next-nearest Na-Na or Si-Na neighbors, as
shown experimentally\cite{meyer:2002}  and
numerically\cite{horbach:2002,sunyer:2003}.
Thus the structure displays some long-range
correlations. One may be tempted to relate these long-range
correlations to the channels discussed above. The validity of this
conclusion, however, is not obvious, since the structure refers to
all ions whereas the channels emerge from the properties of the
specific subset of regions, visited by many different ions.

In this paper we analyze the question to which degree these
conduction channels are predetermined by the structure. This
question is central because it elucidates the role of the network
for long-range transport. Two extreme scenarios are conceivable.
First, following the concepts of Greaves, the network may supply
channels where ion conduction is strongly favored due to the
aggregation of the relevant structural elements (in Greaves work:
non-bridging oxygens) which favor fast dynamics. In this scenario
these relevant structural elements would also display long-range
spatial correlations. Second, the structure might only determine the
local mobility of an ion such that the relevant structural
elements do not display these long-range correlations. Rather the
formation of conduction channels is a simple statistical process
in analogy to the formation of percolation paths in random-energy
or random-barrier lattice models of ion conduction\cite{dyre:2000}.

In previous work we have presented a method how to identify
individual lithium sites which are
basically time-invariant below the glass transition\cite{lammert:2003}.
In particular, we could show that the
continuous trajectories of the lithium ions can be interpreted as
hops between these sites. With this information at hand we can
identify the conduction channels by counting the number of
different ions visiting a specific site during a long MD
simulation. This discretization of the system is somewhat more
adapted to the conduction process than the simple tiling by small
cubes as done by Jund et al \cite{jund:2001}. In particular, we are very
sensitive to the effect of structural elements like
bridging or non-bridging oxygens on the alkali dynamics.

The paper is organized as follows. In Section \ref{sec_tech} we discuss the
technical aspects of our work. Section \ref{sec_result} contains the results
which are finally discussed in Section \ref{sec_dis}.

\section{\label{sec_tech}Technical aspects}

The data for this investigation are based on MD simulations of
lithium silicate $\mathrm(Li_2O)(SiO_2)$. The system containing
1152 atoms was propagated under NVT conditions at 640K with 2 fs
stepsize for 10 ns simulation time. As a basis for investigations
of the cation dynamics we first located the individual sites
available to the lithium ions. For this purpose we have
discretized the system in very small cubes (0.3 \AA) $^3$ and
determined the occupation number in every cube. This is a simple
way to sample the effective potential energy landscape provided by
the silicate network. In order to distinguish regions which indeed
belong to ionic sites and those which only serve as transition
paths between these sites we have used some reasonable cutoff
criterion for the identification of those cubes which are part of
ionic sites. Subsequent cluster analysis of the remaining cubes
finally defines the ionic sites. For our system we have obained 378 
different sites.   Details about the method and about the
properties of the sites are given in an earlier
paper\cite{lammert:2003}.

We characterize the mobility of a site in two different ways.
First, in analogy to the work by Jund et al. the number of {\it
different} ions visiting a site has been determined. It is denoted
\NDIFF{}. Sites for which the value of  \NDIFF{} belongs to the
10\% highest are denoted A-sites. In contrast, those with the 10\%
smallest values of \NDIFF{} are denoted $\bar{A}$-sites. Second,
we determined the average residence time in a site, \TAU{}.
Analoguosly, sites with the 10\% shortest residence times are
denoted B-sites, those with the longest 10\% residence times
$\bar{B}$-sites. In general terms, A- and B-sites indicate
positions in the network where ions are fast, $\bar{A}$- and
$\bar{B}$-sites where ions are slow. In this investigation both
quantities will be used together as different measures of site
mobility.

Furthermore we performed straightforward Monte Carlo simulations
for hopping particles on a 2D square lattice model with 1600 sites
and periodic boundary conditions. We introduced
random site energy disorder and additional barrier disorder among
adjacent sites. Both distributions were chosen as constant with
energies between 0 and 8 $k_B T$. 90\% of all sites were populated
and the hopping particles interacted via simple excluded volume
interaction. The energy disorder was kept constant during the whole run.

\section{\label{sec_result}Results}

\subsection{\label{ssec_gen}General properties of sites}

\begin{figure}
\includegraphics[width=8.6cm,clip]{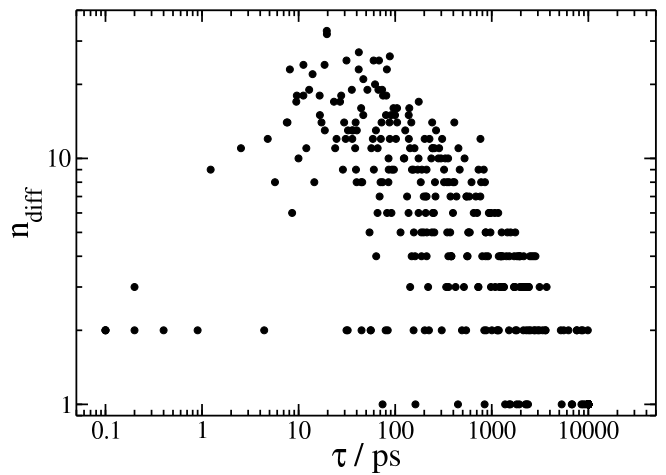}%
\caption{\label{fig_rank} Scatter plot of the number of different
ions \NDIFF{}, vising a site vs. the average residence time of a
site \TAU{}.}
\end{figure}

The two definitions of site mobility are compared in Fig.
\ref{fig_rank}, which shows \NDIFF{} plotted against \TAU{} for
all sites. The highest possible value of \NDIFF{} for a site is
basically given by the number of residences permitted by their average
duration \TAU{}: $\NDIFF{} <= t/\TAU{}$. 
This implies that a sufficiently short average
residence time is a necessary condition for a large number of
different ions visiting a site. A broad range of lower values for
\NDIFF{} is also found, caused by multiple visits of identical
ions. The correlation between \TAU{} and \NDIFF{} is therefore
only weak.

\begin{figure}
\includegraphics[width=8.6cm,clip]{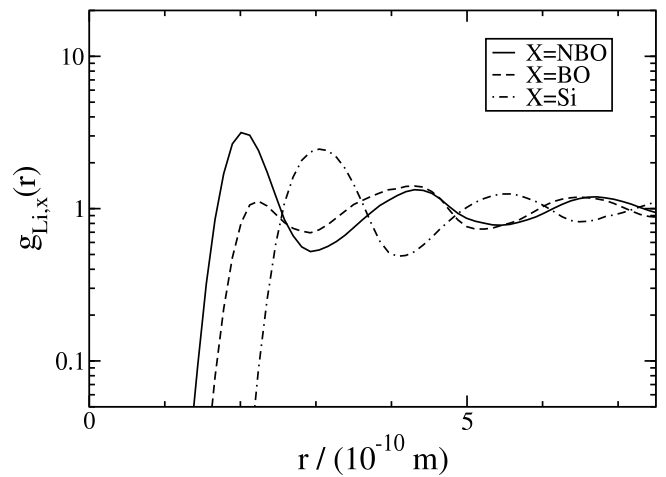}%
\caption{\label{fig_gofr}Mixed g(r) for Li and network species}
\end{figure}

General information on the structural environment of the sites can
be obtained from the radial distribution function g(r) of network
atoms around lithium. It is shown in Fig. \ref{fig_gofr} for
bridging oxygens (BOs), non-bridging oxygens (NBOs), and silicon
atoms. BOs and NBOs have been distinguished according to the
number of silicon atoms within 2.4 \AA{}, the distance of the
first minimum in the g(r) of silicon and oxygen. One finds that
lithium atoms are surrounded by oxygens. Integration up to the
first minimum gives average coordination numbers of 4.0 NBOs and
1.2 BOs. In contrast, if the average number of 5.2 oxygen
neighbors were divided according to the relative numbers of BOs
and NBOs in the system one would obtain 3.3 NBOs and 1.8 BOs. As
expected, NBOs are favored as coordination partners of lithium.
\cite{greaves:1985,kamijo:1996,florian:1996,sunyer:2003:2}.

\subsection{\label{ssec_corr}Spatial correlations of sites}

Now we analyze the question to which degree the mobilities of the
different sites are correlated.  The presence of channels, defined
as regions of the system where many different ions pass would
imply that A-sites (and correspondingly $\bar{A}$-sites) are
spatially correlated over larger distances.  Furthermore, also
possible spatial correlations of B-sites are analyzed.

It can be expected that correlations on the length scale of the
nearest-neighbor distance occur due to trivial reasons. 
For a fast B-site it can be expected that there is a low barrier to an
adjacent site. Accordingly, also the residence time in this
adjacent site can be expected to be small because a fast backjump
to the original site would be possible. This trivial correlation
is limited to next neighbors. If B-sites were arranged into
channels, they must show an additional correlation over longer
distances. The presence of a continuous pathway of B-sites through
the system would result in an increased probability to find other B-sites
among the second neighbors of a given B-site.

\begin{figure}
\includegraphics[width=8.6cm,clip]{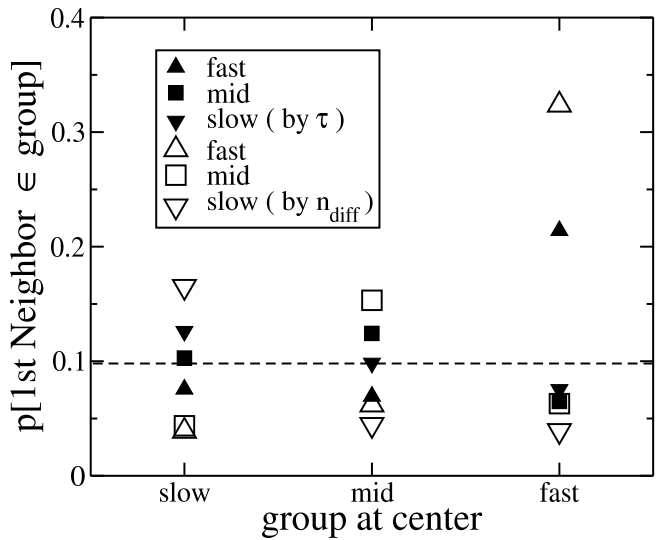}%
\caption{\label{fig_groups}Correlation of mobilities to the next
neighbor shell of sites}
\end{figure}

\begin{figure}
\includegraphics[width=8.6cm,clip]{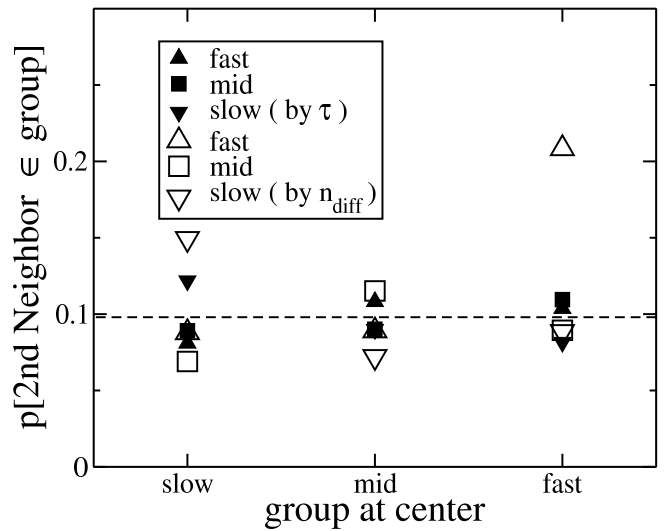}%
\caption{\label{fig_groups2}Correlation of mobilities to the
second next neighbor shell of sites}
\end{figure}

Apart from A- and $\bar{A}$-sites (and analogously B- and
$\bar{B}$-sites) we also selected a subset of 10\% of  
sites with intermediate \TAU{} and \NDIFF{}, respectively. Each of the
three groups
contained 37 sites, i.e. ca. 10\% of the total number. Because of
the low number of sites we did not compute their complete pair
distribution functions g(r). Instead we counted directly the
number of neighbors from each of the three groups around a given
central site. The limits of the first and second neighbor shell
were taken from the minima in the g(r) of all sites at 4.2 \AA{}
and 6.8 \AA{}. The results for slow, intermediate and fast central
sites are compared in Fig. \ref{fig_groups} and Fig.
\ref{fig_groups2} for the first and for the second shell
respectively.

All values are given as a fraction of the total number of
neighbors. The statistical value of $37/378 = 9.8 \%$ for the
contribution of each group is shown by the dashed lines.
Generally, there is a clear tendency that a site will be
surrounded by first neighbors of similar mobility. The strongest
correlation is found among the fast sites. While the findings are
similar for both \TAU{} and \NDIFF{}, they are more pronounced in
the latter case. For the second shell of neighbors, the
correlations almost vanish for the residence time \TAU{}, i.e.
B-sites (and $\bar{B}$-sites) are spatially {\it not} correlated.
In contrast, as already reported by Jund et al. A-sites are
significantly correlated beyond the nearest-neighbor shell. The
correlation for $\bar{A}$-sites is much weaker, but still present.

\subsection{\label{ssec_net}Spatial correlations of relevant network properties}

Next we turn to the question in which sense the conduction
channels are predetermined by the structure. To answer this
question we have to proceed in two steps. First, we have to check
which property of the network significantly favors highly mobile
sites. As will be shown below it is the presence of bridging
oxygens (BOs) which displays the strongest correlation with the
mobility of a nearby site. Second, possible long-range spatial
correlations of this network property (here: occurrence of BOs)
have to be analyzed. More specifically we check whether sites with
a high/low number of BOs are spatially correlated. Again we have
to be aware that trivial nearest-neighbor correlations will be
present. If some site displays an increased number of BOs as
compared to the average site it can be expected that a small
number of neighbor sites shows the same correlation because one BO
typically belongs to 2 lithium ions.

\begin{figure}
\includegraphics[width=8.6cm,clip]{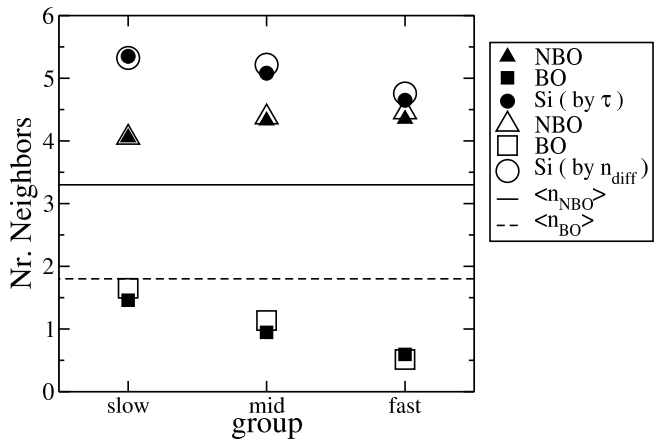}%
\caption{\label{fig_net}Network species as next neighbors of
sites}
\end{figure}

Using the sets of sites already defined above, we first 
count the number of BOs, NBOs and Si atoms up to the 
1st minimum distances of 2.9\AA{}, 3.0\AA{}, and 4.1 \AA{} respectively.
The results are shown in Fig. \ref{fig_net}. As expected from
the above-mentioned analysis of coordination numbers,
sites are on average surrounded by
$\approx$ 4 NBOs, while only 0.5 to 1.5 BOs are found.
For comparison,
lines are shown for 1.8 BOs and 3.3 NBOs, i.e. the statistical
values (see above). Interestingly, the strongest influence on the
mobility of a site is shown for the number of coordinated BOs,
which decreases by $\approx 1$ from slow to fast sites. The number
of silicon neighbors shows only a weak decrease in the same
direction. For NBOs the correlation is reverse, but much weaker
even on an absolute scale. The results are very similar when using
\TAU{} or \NDIFF{} to determine the mobility of the sites.

\begin{figure}
\includegraphics[width=8.6cm,clip]{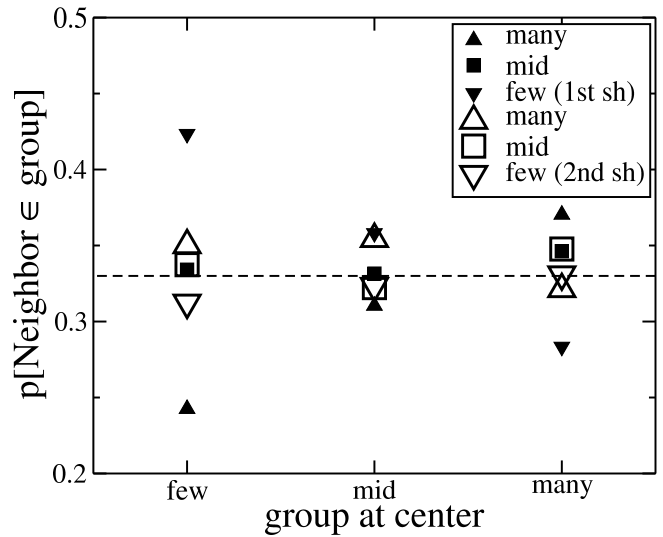}
\caption{\label{fig_bo}Correlations in the number of neighboring BOs of sites}
\end{figure}

This result implies that in most cases a single BO in the
nearest-neighbor shell is sufficient to immobilize ions on the
sites around the BO.  A possible underlying structure in the
network that could determine the layout of conduction channels
should therefore be related to the distribution of BOs. Similarly
to the method described above, we sorted all sites into three
groups according to the number of BOs in their 1st shell closer
than 2.9 \AA{}. Then the fraction of neighbors from each of these
groups was counted. The results are given in Fig. \ref{fig_bo}.
The dashed line again marks the statistical value, which is 0.33
in this case, corresponding to groups of one third of all sites.
The expected trivial correlation among nearest neighbors is, of
course, found. Most important, no correlation to the second
nearest neighbor shell is found. We checked that also the NBOs do
not display any correlations beyond the nearest neighbor shell.
The lack of spatial correlations beyond the nearest neighbors
strongly suggests that even the distribution of bridging oxygen atoms
as the most relevant network property does not determine
complete conduction channels in the system.

\section{\label{sec_dis}Discussion}

\begin{figure}
\includegraphics[width=4.2cm,clip]{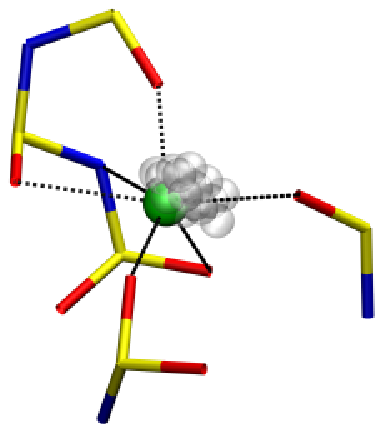}%
\includegraphics[width=4.2cm,clip]{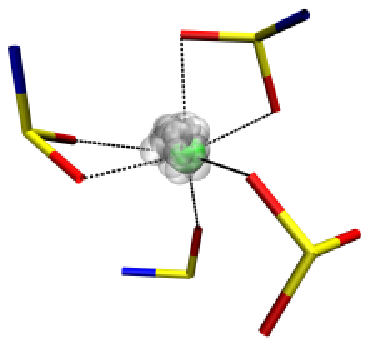}%
\caption{\label{fig_local}Local structure around a slow site (left) and a fast site (created with the software vmd\cite{vmd})}
\end{figure}

The presence of only one BO drastically slows down the ionic
transport through nearby sites. An indication how this slowdown is
effected can be obtained from the two examples shown in
Fig.\ref{fig_local}. Here the local network structure is shown around
one slow site (left) and one fast site. Cations are shown at the
center as solid balls.
% The network atoms are represented by translucent spheres, and also
% by tubes giving the connectivity.
The surrounding silicate network is shown as tubes, .
 The site as determined by us\cite{lammert:2003} is represented
 by the translucent cloud surrounding the cation.
Connections are drawn
between the lithium ion and all oxygen atoms within the first
neighbor shell. For the fast site, all six coordinating
oxygen atoms are NBOs, which are bound to four different silicon
atoms. For the slow site there is one BO among the neighbors. This
BO is part of a silica chain segment that is also carrying three
of the NBOs coordinating the cation. These NBOs are bound to three
silicon atoms on different sides of the BO, with an additional BO
inserted on one side in between. In total, a chain segment of 5
atoms is linked to the cation via four coordinating oxygen atoms.
The chain is thus closely wrapped around a large part of the site.
This situation suggests that BOs introduce high transport barriers
for nearby sites because the silica chains joined by them block a
large fraction of possible jump pathways.

The pathways through the system used for effective ionic transport
follow continuous channels of sites visited by many ions. The passage
of a large number of different ions through a specific site 
depends on two conditions. (i) A comparatively low average
residence time of the sites is necessary but not sufficient to
enable the passage of many ions. (ii) The presence of coordinating
BOs strongly decreases the number of ions reaching a site,
probably by introducing a local barrier. Both the distribution of
sites with low \TAU{} and that of BOs in the structure show no
long range order comparable to that found in the pathways traced
by sites with high \NDIFF{}. They can therefore only impose
constraints on the layout of these pathways, but not determine
them. Therefore the emergence of conduction channels does not seem
to require preformed channels provided by the network structure.

\begin{figure}
\includegraphics[width=8.6cm,clip]{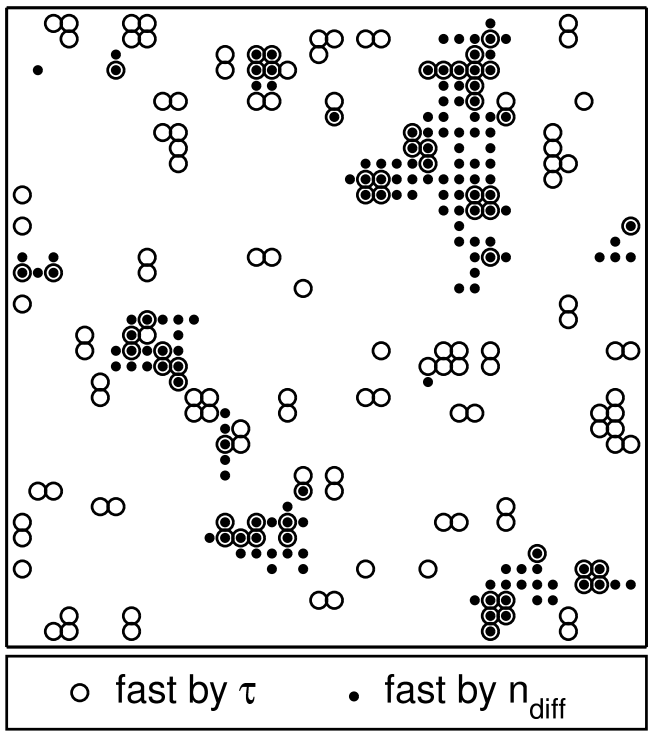}%
\caption{\label{fig_chan}Spatial correlations in a simple lattice model}
\end{figure}

Now we show that on a qualitative level similar effects can
be observed for the simple
lattice model, introduced in Section \ref{sec_tech}.
After performing long Monte
Carlo simulations we determined the A- and B-sites similarly to
our procedure for the silicate system. They are marked by filled 
dots and open circles, respectively. Interestingly, the three
main observations from the silicate system are reproduced: (i) By
definition the structure (here: the local energetic parameters)
does not display correlations beyond the nearest neighbor shell.
(ii) The B-sites are not spatially correlated beyond the
nearest-neighbor shell. Actually, one clearly sees that sites with
short residence time typically occur as pairs, because of the trivial
correlations discussed above. 
(iii) The A-sites are spatially
correlated, i.e. they are concentrated in a few larger blobs.
These regions contain some of the groups of sites with low \TAU{},
but also some sites with higher residence times. Correspondingly,
some sites with low \TAU{} are not incorporated into regions with high \NDIFF{}.
Sites visited by many different ions
can of course not be completely isolated. The ions have to move away
through other sites in the neighborhood to allow the following ions to
enter. In our lattice model the regions best suited for this
process are located by chance. The ions benefit from sites with
very low residence times, they must reach them via other sites
that are also not too slow, and they must avoid high barriers
distributed through the lattice. Interestingly, the pathways are
somewhat extended in both dimensions. This suggests that regions
where many different particles may diffuse 
are extended in all directions. Actually, the channels, as
obtained from the molecular dynamics simulations\cite{jund:2001} show
this property as well.

%\begin{figure}
%\includegraphics[width=8.6cm,clip]{fig8.eps}%
%\caption{\label{fig_chan}Channels in a simple lattice model}
%\end{figure}

In summary, we have analyzed the nature of the conduction pathways
in lithium silicate as well as a simple random lattice model. In
both cases we do find spatial correlations in regions of high mobility.
These structures do not
seem to be supplemented by corresponding spatial correlations of
network properties, following from the lack of long-range
correlations of sites with BOs. Thus we would like to suggest that
the formation of conduction channels is rather a statistical
selection process in analogy to the formation of percolation paths
in simple lattice models. % like random energy or random barrier models.

\begin{acknowledgments}
We thank M. Korth and H. Krieg for the help with the MC-simulations.
Furthermore we acknowledge very helpful conversations with P. Jund.
H. L. also acknowledges the support by the Fonds der Chemischen Industrie.
\end{acknowledgments}

%\bibliography{reflist}

\end{document}